\documentclass[conference, 10pt]{IEEEtran}
\IEEEoverridecommandlockouts

\usepackage{cite}
\usepackage{amsmath,amssymb,amsfonts}
\usepackage{algorithmic}
\usepackage{graphicx}
\usepackage{textcomp}
\usepackage{xcolor}
\usepackage{orcidlink}
\usepackage{listings}
\usepackage[font={small}]{caption} 
\usepackage{subcaption}
\usepackage[utf8]{inputenc}
\lstdefinestyle{base}{
  language=C++,
  emptylines=1,
  breaklines=true,
  basicstyle=\ttfamily\small\color{black},
  moredelim=**[is][\color{red}]{@}{@},
  moredelim=**[is][\color{blue}]{|}{|},
}

\newcommand{\nbX}{non-blocking}
\newcommand{\NbX}{Non-blocking}
\newcommand{\nb}{\nbX~}
\newcommand{\Nb}{\NbX~}

\newcommand{\queueX}{\textit{Queue}}

\newcommand{\queue}{\queueX~}

\newcommand{\namespace}{\texttt{MPI\_}}

\newcommand{\MPISX}[1]{\namespace\texttt{#1}}
\newcommand{\MPIS}[1]{\MPISX{#1}~}

\begin{document}

\title{Co-Design and Evaluation of a CPU-Free MPI GPU Communication Abstraction and Implementation \thanks{Notice: This manuscript includes authors employed by UT-Battelle, LLC, under contract DE-AC05-00OR22725 with the US Department of Energy (DOE). The US government retains and the publisher, by accepting the article for publication, acknowledges that the US government retains a nonexclusive, paid-up, irrevocable, worldwide license to publish or reproduce the published form of this manuscript, or allow others to do so, for US government purposes. DOE will provide public access to these results of federally sponsored research in accordance with the DOE Public Access Plan (\url{https://www.energy.gov/doe-public-access-plan}).}}

\makeatletter
\newcommand{\linebreakand}{%
  \end{@IEEEauthorhalign}
  \hfill\mbox{}\par
  \mbox{}\hfill\begin{@IEEEauthorhalign}
}
\makeatother

\author{
\IEEEauthorblockN{Patrick G. Bridges \orcidlink{0000-0003-4801-0390}}
\IEEEauthorblockA{University of New Mexico\\
Albuquerque, New Mexico, USA \\
patrickb@unm.edu
}
\and
\IEEEauthorblockN{Derek Schafer \orcidlink{0000-0001-8438-5144}}
\IEEEauthorblockA{University of New Mexico\\
Albuquerque, New Mexico, USA \\
dschafer1@unm.edu
}
\and
\IEEEauthorblockN{Jack Lange \orcidlink{0000-0003-0616-7437}}
\IEEEauthorblockA{Oak Ridge National Laboratory\\
Oak Ridge, Tennessee, USA \\
langejr@ornl.gov}
\linebreakand
\IEEEauthorblockN{James B.~White III \orcidlink{0009-0005-2186-075X}}
\IEEEauthorblockA{Oak Ridge National Laboratory\\
Oak Ridge, Tennessee, USA \\
whiteiiijb@ornl.gov}
\and
\IEEEauthorblockN{Anthony Skjellum \orcidlink{0000-0001-5252-6600}}
\IEEEauthorblockA{Tennessee Tech University\\
Cookeville, Tennessee, USA\\
askjellum@tntech.edu}
\and
\IEEEauthorblockN{Evan Suggs \orcidlink{0000-0002-8210-8992}}
\IEEEauthorblockA{Tennessee Tech University\\
Cookeville, Tennessee, USA\\
esuggs@tntech.edu}
\linebreakand
\IEEEauthorblockN{Thomas Hines \orcidlink{0000-0001-9675-0399}}
\IEEEauthorblockA{Tennessee Tech University\\
Cookeville, Tennessee, USA\\
tmhines42@tntech.edu}
\and
\IEEEauthorblockN{Purushotham Bangalore \orcidlink{0000-0002-1098-9997}}
\IEEEauthorblockA{University of Alabama \\
Tuscaloosa, Alabama, USA \\
pvbangalore@ua.edu}
\and
\IEEEauthorblockN{Matthew G.~F.~Dosanjh \orcidlink{0000-0001-5141-9176}}
\IEEEauthorblockA{Center for Computing Research \\
Sandia National Laboratories\\
Albuquerque, New Mexico, USA \\
mdosanj@sandia.gov}
\and
\IEEEauthorblockN{Whit Schonbein \orcidlink{0000-0003-4955-2984}}
\IEEEauthorblockA{Center for Computing Research\\
Sandia National Laboratories\\
Albuquerque, New Mexico, USA\\
wwschon@sandia.gov}
}

\maketitle

\begin{abstract}
Removing the CPU from the communication fast path is essential to 
efficient GPU-based ML and HPC application performance. However, 
existing GPU communication APIs either continue to rely on the 
CPU for communication or rely on APIs that place significant 
synchronization burdens on programmers. In this paper we describe 
the design, implementation, and evaluation of an MPI-based GPU 
communication API enabling easy-to-use, high-performance, 
CPU-free communication. This API builds on previously proposed 
MPI extensions and leverages HPE Slingshot 11 network card 
capabilities. We demonstrate the utility and performance 
of the API by showing how the API naturally enables CPU-free 
gather/scatter halo exchange communication primitives in the 
Cabana/Kokkos performance portability framework, and through a 
performance comparison with Cray MPICH on the  
Frontier and Tuolumne supercomputers. Results from this evaluation show up 
to a 50\% reduction in medium message latency in simple GPU ping-pong 
exchanges and a 28\% speedup improvement when strong scaling a 
halo-exchange benchmark to 8,192 GPUs of the Frontier 
supercomputer.



\end{abstract}


\section{Introduction}
\label{sec:intro}

Modern HPC systems include sophisticated network capabilities to support communication using GPU devices. For example, recent GPU and network architectures include features to support 
GPU-initiated communication~\cite{agostini_gpudirect_2018,openfabrics_interfaces_working_group_fi_cxi_nodate,hamidouche_gpu-initiated_2025}.
These features are designed to enable efficient direct GPU-to-NIC (network interface card) communication and to remove the CPU from the communication fast path; a recent survey of GPU communication techniques termed this overarching goal \emph{CPU-free communication}~\cite{unat_landscape_2024}.

Traditional approaches to communicating GPU payloads require CPU/GPU synchronization. GPU-initiated, CPU-free communication seeks to provide high-performance communication APIs that do not require this CPU/GPU synchronization. This optimization may reduce communication latency and increase small-message bandwidth, while improving the performance of strong scaling workloads whose performance is dominated by communication latency. 

Currently proposed CPU-free communication interfaces and implementations often significantly reduce the expressiveness of the communication API. For example, some remove support for features commonly used in HPC applications, such as message matching~\cite{agostini_gpudirect_2018,nvidia_nvshmem_2025,nvidia_nccl_2025,namashivayam_exploring_2023} and two-sided data movement~\cite{nvidia_nvshmem_2025}, from the available APIs to achieve the benefits of CPU-free communication; this has limited their usability in HPC applications and performance portability frameworks. More general-purpose GPU communication APIs, including most proposed GPU-triggered MPI APIs~\cite{Bridges:2025:Understanding} and NVIDIA NCCL~\cite{nvidia_nccl_2025}, still generally use the CPU on the communication fast path\footnote{NCCL release 2.28.3 recently added experimental support for what it terms GPU-Initiated Networking~\cite{hamidouche_gpu-initiated_2025}.}.


This paper describes the design, implementation, and evaluation of MPI extensions that provide CPU-free, two-sided MPI GPU communication.
Specifically, after presenting relevant background information (Section~\ref{sec:background}), this paper describes in detail the following research contributions:
\begin{itemize}
\item The design of a new GPU communication API that can support CPU-free communication and mostly preserves familiar MPI two-sided communication abstractions (Section~\ref{sec:api}).
\item The demonstration of the usage of this API to create revised versions of halo exchange objects used by both benchmarks and production applications written for the Cabana performance portability framework (Section~\ref{sec:example}).
\item An implementation of this API for HPE Slingshot 11 systems~\cite{de_sensi_-depth_2020} that leverages the deferred work queue and counter features of the \texttt{libfabric} API~\cite{grun_brief_2015} and the HPE \texttt{libfabric} CXI provider~\cite{openfabrics_interfaces_working_group_fi_cxi_nodate} to provide full CPU-free MPI point-to-point communication (Section~\ref{sec:implementation}).
\item The evaluation of the impact of these abstractions on MPI GPU ping-pong performance and the strong scaling performance of HPC benchmarks, which demonstrates up to 50\% reduction in GPU ping-pong latency and 20\% or more improved strong scaling in Cabana halo-exchange benchmarks, when running on 1,024 nodes with 8,192 GPUs of the ORNL Frontier supercomputer (Section~\ref{sec:evaluation}).
\end{itemize}
In addition, this paper also discusses related work (Section~\ref{sec:related}) and directions for future work (Section~\ref{sec:future}) relevant to these research contributions, prior to presenting conclusions (Section~\ref{sec:conclusions}).

\section{GPU Communication Background}
\label{sec:background}

Communication on modern GPU systems typically involves both GPU and CPU execution, with significant latency overheads. Even the simple sequence of receiving data, running a kernel to compute using the received data, and then sending a message based on the results of that computation involves complex GPU, CPU, and NIC interactions. 
Our research builds on multiple approaches to reduce these overheads, many of them leveraging features of modern communication devices. The remainder of this section provides background on GPU communication costs, the approaches to mitigating or eliminating them our research builds on, the \texttt{libfabric} features and extensions implemented on HPE Slingshot NICs that we leverage to reduce these overheads, and performance data on the impact of different HPC GPU communication approaches on latency.

\subsection{Overheads in GPU-Aware Communication Exchanges}

Standard GPU-aware MPI communication incurs significant latency overheads, as shown in Fig.~\ref{fig:mpi-exchange}. 
\begin{figure}
\begin{lstlisting}[style=base]
MPI_Recv(messageIn, src, length);
parallel_kernel<x, y, stream>(problem, 
    messageIn, messageOut); // Kernel Launch
gpuStreamSynchronize(stream); // Memory Barrier
MPI_Send(messageOut, dest, 
    length); // RTS/CTS and Matching
\end{lstlisting}
\caption{Simplified CPU-driven GPU computation and MPI communication. Most current applications use this CPU-based approach to communicating data to and from GPU computations.}
\label{fig:mpi-exchange}
\end{figure} In addition to network communication costs, GPU-aware two-sided MPI communication incurs the following costs:
\begin{LaTeXdescription}
    \item[Kernel Launch Latency:] Launching a parallel kernel from the CPU typically takes on the order of a microsecond.
    \item[GPU Memory Barrier:] GPUs include significant caches, and even when CPU and GPU memories are coherent, communicating data from GPU memory can require a memory barrier (the main cost of a stream synchronization) that takes on the order of a microsecond.
    \item[Messaging Setup Costs:] Traditional two-sided MPI communication, which ensures that the receive buffer is ready, for example using a Ready-to-Send/Clear-To-Send (RTS/CTS) protocol, and handles MPI matching semantics, can take several microseconds depending on the exact details of the messaging implementation.
\end{LaTeXdescription}
As a result, GPU communication can easily increase latency by 2-4 microseconds. For 16KB message (e.g., an edge in a halo exchange) where network latency is on the order of 5 to 10 microseconds, this could increase end-to-end latency by 20-80\%.


\subsection{Reducing GPU Communication Overheads}

A wide range of approaches have been proposed for reducing the GPU communication overheads described above, which has been summarized by Unat et al\hbox{.}~\cite{unat_landscape_2024}. The most relevant of these, which inspired techniques used in this paper, are the following: 
\begin{LaTeXdescription}
    \item[GPU-initiated CPU-driven communication:] In this approach, GPU kernels interact with CPU communication threads that poll memory, waiting for GPU kernel completion before initiating communication. This reduces kernel launch and stream synchronization latencies, and is typified by the NVIDIA Collective Communication Library~\cite{nvidia_nccl_2025}, the \verb|MPIX_Stream| extension~\cite{zhou_mpix_2022} to the \texttt{MPICH} communication library, and HPE's original GPU-based communication primitives~\cite{namashivayam_exploring_2022}.
    \item[CPU-free one-sided communication:] In this approach, GPU code, whether stream or kernel triggered, uses a restricted communication interface that eliminates the complex synchronization logic used to match messages, pauses the sender to ensure buffer readiness, and notifies the receiver of data availability. This enables implementations to eliminate kernel launch, memory barrier, and messaging setup costs described above, at the cost of increased programmer synchronization management. NVIDIA's NVSHMEM library is the canonical example of this approach~\cite{nvidia_nvshmem_2025} and HPE has also published a paper describing a CPU-free MPI one-sided interface~\cite{namashivayam_exploring_2023}. 
    \item[CPU-free two-sided communication:] These approaches restrict or modify the semantics of  two-sided semantics to enable CPU-free communication. In the first such approach, NVIDIA researchers designed the \texttt{libmp} two-sided communication library~\cite{agostini_gpudirect_2018}. While this library provided an MPI-like interface, it did not support MPI matching and required that the sender separately guarantee that the receiver was ready to receive sent data. Recent work with MPI partitioned communication moved matching off the two-sided communication critical path and introduced a new MPI call, \verb|MPIX_Pbuf_prepare|, to guarantee receiver buffer readiness; this call required CPU involvement prior to each GPU communication cycle~\cite{temucin_design_2024} (however, this function is not yet adopted formally as of the MPI 5.0 standard~\cite{mpi50}). 
\end{LaTeXdescription}

\subsection{Slingshot 11/\texttt{libfabric} GPU Communication Features}
\label{sec:background:cxi}
HPE Slingshot 11 network devices implement the general Open Fabrics Interconnect interface through the open-source CXI provider~\cite{openfabrics_interfaces_working_group_fi_cxi_nodate}. This provider includes implementations of the OFI triggered communication operations originally designed for the Portals communication API~\cite{underwood_enabling_2011,portals_43}. The \texttt{libfabric} CXI provider for these devices also includes API extensions to access these features from GPUs, which we describe in this section. They provide the foundation for CPU-free GPU-triggered communication that we leverage in this paper.

\texttt{libfabric} defines two features to enable construction of triggered communication: (1) \emph{triggering counters} that count network device events and can be configured to increment either upon the occurrence of specific network events (e.g., the local completion of a local operation such as a send, atomic operation, or a remote write from a remote network device) or by explicit manipulation by the host, and (2) \emph{deferred work queue entries} (DWQs) that specify \texttt{libfabric} communication operations\footnote{\texttt{libfabric} deferred work queue entries currently support only point-to-point communication operations, not hardware collective operations.} whose execution is delayed until an associated triggering counter reaches a specified integer value.

HPE's CXI provider includes key extensions to the counter API calls that enable the GPU to trigger counters. First, the CXI provider can give the user an address of the counter that is in a memory-mapped I/O region, which the user can then register with the GPU to be used in a kernel or a  GPU stream memory operations (e.g., \texttt{hipStreamWriteValue64}). Second, the CXI provider offers registration for a custom writeback buffer for the counter. Note, however, reading CXI counter values requires CPU progress to update host writeback buffers. As a result, GPU memory operations cannot directly read CXI counters without CPU involvement.


HPE has previously demonstrated the use of these features to support CPU-free GPU-triggered one-sided MPI communication~\cite{namashivayam_exploring_2023}. Importantly, this paper described an approach to enable GPUs to poll for completion of remote writes without CPU intervention by having the \emph{sender} trigger an \texttt{fi\_atomic} write to a host memory location on the receiver when the \texttt{fi\_write} completes on the sender. The receiving GPU, which cannot poll CXI counters directly without CPU assistance, can instead poll this memory location to wait for receive completion. As described in Section~\ref{sec:implementation}, we use a variation of this approach where an \texttt{fi\_atomic} is triggered by an \texttt{FI\_REMOTE\_WRITE} memory region counter on the \emph{receiver}, allowing the GPU to wait for receive completion without an extra network round trip. 



\section{Proposed MPI GPU Triggering API}
\label{sec:api}


Because of the importance of two-sided communication to a wide range of HPC applications, 
we set out to design an API that supports as many of the 
two-sided and collective MPI communication modes as possible. In addition, we sought to design an API that would support CPU-free GPU control flow and data movement on the communication fast path of HPE Slingshot 11 network devices. In the remainder of this section, we provide an overview of our design and the rationale behind our design choices, and we then give a detailed semantics of the key elements of this API.

\subsection{Overview}

Our first design decision for a new MPI GPU API was to \textbf{use existing MPI persistent operations}, specifically persistent point-to-point and collective operations, whenever possible \cite{mpi50}. The existing persistent operations are attractive for GPU communication for two reasons: (1) they cover almost all of the commonly used MPI semantics, reducing the number of new abstractions or optimizations that need to be added to MPI, and (2) they explicitly separate setup operations involved in creating a request from the use of that request on the communication fast path. The latter is particularly important because it allows our \MPISX{Match} function described below to remove communication operations with complex semantics such as message matching and exchange of RMA keys from the GPU-NIC fast path, where such operations are generally infeasible~\cite{klenk2017relaxations}.

Next, we \textbf{added an MPI progress engine abstraction, the \MPISX{Queue}, that ties persistent operation communication to GPU execution}.
In this model, \emph{start} and \emph{wait} operations are enqueued to an MPI queue that is associated with a GPU stream, similar to the (non-persistent) approach first suggested by HPE researchers~\cite{namashivayam_exploring_2022}. 
We augmented that proposal by also adopting the concurrency semantics proposed by MPICH researchers as part of the \texttt{MPIX\_Stream} proposal, though not the more complex stream communicator elements of that proposal~\cite{zhou_mpix_2022}.

Third, we \textbf{added a set of \MPISX{Match} operations that can be applied to existing persistent operations}. This change enables one-sided data movement for two-sided and collective persistent operations. This is not currently possible for MPI persistent point-to-point operations, with the exception of MPI partitioned point-to-point operations~\cite{mpi50}. This also address OFI deferred work queues semantics not supporting the deferred activation of \texttt{fi\_recv} operations~\cite{grun_brief_2015}. 

Finally, we \textbf{retained required MPI buffer readiness guarantees on the GPU-NIC fast path}. Deferring data movement until the receiver indicates the remote buffer is ready is one of the features that makes two-sided communication more convenient than one-sided communication. Previous MPI GPU approaches added CPU involvement to the GPU communication fast path to ensure readiness (e.g., \texttt{MPI\_Pbuf\_Prepare}, a proposed addition to partitioned communication mentioned above that is not currently in the MPI standard \cite{message_passing_interface_forum_mpi_2025}). However, we determined that such calls were (1) unnecessary if we carefully leverage MPI persistent operations and OFI deferred work queue/triggered operation semantics, and (2) were undesirable if the application could separately guarantee and indicate buffer readiness (e.g., through the use of MPI ready send operations).


\subsection{\MPISX{Queue}: Ordering Progress with External Execution} 
To represent an ordered set of communication tasks to execute on a GPU, we define an \MPISX{Queue} object. Proposed APIs related to general management of the \queue object itself can be found in in Table~\ref{tbl:queue_init}. These APIs are similar to other creation and freeing functions found within the MPI standard. 
The remainder of this section explains how to use the proposed \queue objects. 

\begin{table}[ht]
\centering
\caption{Queue Object Management APIs}
\begin{tabular}{|l|l|l|l|}
    \hline 
     \textbf{API} & \textbf{Description} & \textbf{Input(s)} & \textbf{Output} \\
    \hline 
    \textit{Queue\_init} & Create a \queue & Type of \queue, & \queue \\
    & object & Address of related & \\
    & & GPU stream & \\
    \hline
    \textit{Queue\_free} & Free a \queue   & \queue to free &  \\
    & object and any  & & \\
    & resources & &\\
    \hline
\end{tabular}
\label{tbl:queue_init}
\end{table}

\subsection{\MPISX{Match}: Enabling Stream Triggering of Legacy Persistent Operations}

In MPI, blocking, \nbX, and persistent point-to-point calls need to be matched before data transfer can occur. For persistent calls, this matching occurs and adds complexity on every use. We mitigate this by requiring stream-triggered operations to have been fully matched---the implementation must have already done tag matching and should not need to negotiate with its communication partner how and where to put the data being exchanged. While not a new concept to MPI, we propose an API that provides the user with more explicit control of when this action happens, as well as the means to check on the status of matching. Table~\ref{tbl:matching} outlines the set of APIs we are proposing.

\MPISX{Match} accepts an MPI request representing an inactive persistent request and returns a permanently matched equivalent. This means that once a request is processed by \MPISX{Match}, it will have a determined pair request on the remote process, which will be permanent until the request is freed through \texttt{MPI\_Request\_free}. This eliminates much of the complexity of point-to-point, allowing for various optimizations and enabling stream triggering. 

Currently in MPI, any variant of \texttt{MPI\_Send} may match with any style of \texttt{MPI\_Recv}. This property of matching means that persistent communication may match with non-persistent communication, and thus persistent communication requests must match every time they are started. 
To ensure a persistent match, a persistent request using our matching APIs may only match to other persistent operations. \Nb requests cannot be matched with this API, either as direct inputs or as the matched operation on the other end of a matching call. Generalized requests are also considered as erroneous input. If a persistent request goes through the normal MPI calls and does not use any of our proposed APIs, then there is no change in behavior.

In addition to a blocking match function, we also provide a \nb version. This function will return an \texttt{MPI\_Request} that the user can wait and test on, similar to other \nb MPI communications. These matching requests are \nbX, non-persistent, and cannot be canceled. When doing a \nb matchall of multiple requests, only a single request is returned from the function. This matching request will be considered completed when all of the input requests have been matched.

Lastly, we also provide an API to test whether the original communication request(s) 
has/have been matched. The function does nothing to the matching request regardless of the state of the match; the matching request must still be completed by a regular MPI function (i.e.,~\texttt{MPI\_Test} or \texttt{MPI\_Wait}).

\begin{table}[b]
\centering
\caption{Proposed Matching APIs}
\begin{tabular}{|l|l|l|l|}
    \hline 
     \textbf{API} & \textbf{Description} & \textbf{Input(s)} & \textbf{Output} \\
    \hline 
    \textit{Match(all)} & Block until given & Request(s) &  \\
    &  Request(s) are matched &(array size)  & \\
    \hline
    \textit{Imatch(all)} & Start the match & Request(s) & New Request \\
    & for Request(s) & (array size) & (array) \\
    \hline
    \textit{Is\_matched} & Query the match  & Request & Match result \\
    & status of a Request & & \\
    \hline
\end{tabular}
\label{tbl:matching}
\end{table}

\subsection{Concurrency Management using \texttt{MPI\_Queue}}

\begin{table}[ht]
\centering
\caption{Proposed Enqueueing APIs}
\begin{tabular}{|l|l|l|l|}
    \hline 
     \textbf{API} & \textbf{Description} & \textbf{Input(s)} \\
    \hline 
    \textit{Enqueue\_start(all)} & Enqueue the start of a & Queue  \\
     & Requests(s) on a \queue & Request(s) \\
     & & (array size) \\
     \hline
    \textit{Enqueue\_wait(all)} & Enqueue a wait that will  & Queue  \\
     & block a \queue until  & Request(s) \\
     & Request(s) are completed & (array size) \\
     \hline
    \textit{Queue\_wait} & Block the caller until & Queue  \\
    &  a \queue is empty & \\
    \hline
\end{tabular}
\label{tbl:enqueue}
\end{table}

Once an \MPISX{Queue} object has been created, users may begin adding matched persistent requests to a \queue using the APIs shown in Table~\ref{tbl:enqueue}. It is erroneous to add \nb requests or generalized requests to a \queueX. If the user enqueues a request that has not yet been matched, the enqueue function will return an error, and the request will not be enqueued. If this happens in a function where multiple requests are enqueued, no request will be enqueued. 
In order for a request to be permitted to enqueue its start, it must be a persistent request, it must be matched, and any previous starts enqueued for the request must have a corresponding wait.

After a request has been enqueued, the host thread is not permitted to call most normal MPI operations on it (e.g. \texttt{MPI\_Wait}, \texttt{MPI\_Cancel}, etc.). In addition, a request that has started on one \queue may not be waited on in a different \queueX. 


Finally, we provide a function to block the CPU thread until all actions (communication or otherwise) in the \queue have completed. This call can be thought of as doing, at a minimum, the relevant synchronization method (e.g., \texttt{cudaStreamSynchronize}), but an implementation may also do other progression related activities as well. Once the user knows that an enqueued request has been completed, the programmer is allowed to call \texttt{MPI\_Request\_free} to free the request.


\section{Prototyping Halo Exchanges with the Revised API}
\label{sec:example}

\begin{figure}
\begin{lstlisting}[style=base]
StreamHalo(ExecutionSpace exec_space)
{
  ... 
  _my_stream = findStream(exec_space); 
  MPI_Queue_init(&_my_queue, CXI, 
      &_my_stream);
  // Create Send/Receive Requests
  for (int n = 0; n < num_n; ++n)
  {   
    MPI_Recv_init(_ghost_buffers[n].data(),
      _ghost_buffers[n].size(), MPI_BYTE,
      _neighbor_ranks[n], _receive_tags[n],
      _comm, &_requests[n]);
    MPI_Send_init(_owned_buffers[n].data(),
      _owned_buffers[n].size(), MPI_BYTE, 
      _neighbor_ranks[n], _send_tags[n], 
      _comm, &_requests[num_n + n]);
  }
  // Match send and receive requests
  MPI_Matchall(&_requests[n]);
  ...
}
\end{lstlisting}
\caption{Simplified version of MPI code to initialize a halo gather operation in the \texttt{StreamHalo} object. 
}
\label{fig:send-recv-init}
\end{figure}

\begin{figure}
\begin{lstlisting}[style=base]
template <class... ArrayTypes>
void enqueueGather(
  const ArrayTypes&... arrays)
{
  // Enqueue starting the receives
  MPI_Enqueue_startall( _my_queue, 
    num_n, &_requests[0] );
  // Enqueue packing the data to send
  enqueuePackBuffers( _owned_buffers,
    _owned_steering, arrays.view()...);
        
  // Enqueue sends and waiting for 
  // communication to complete
  MPI_Enqueue_startall( _my_queue, 
    num_n, &_requests[num_n] );
  MPI_Enqueue_waitall( _my_queue );

  // Enqueue unpacking receive buffers
  enqueueUnpackBuffers(_ghost_buffers,
    _ghost_steering, arrays.view()...);
}
\end{lstlisting}
\caption{Simplified version of MPI code to enqueue a ghost cell gather operation in the \texttt{StreamHalo} object. 
}
\label{fig:enqueuegather}
\end{figure}

The StreamHalo class discussed in this paper derives from the Halo class in Cabana's grid 
package~\cite{slattery_cabana_2022}.
This paper covers two version of StreamHalo, a default MPI-based StreamHalo class and a version enhanced to use the stream-triggering API described in Section~\ref{sec:api}. 
Both versions implement two main methods: 
\texttt{enqueueScatter} and \texttt{enqueueGather};  the former uses standard MPI \texttt{Isend} and \texttt{Irecv}, while the latter uses our prototype stream-triggering APIs.

Fig.~\ref{fig:send-recv-init} shows part of the constructor for the version of the \texttt{StreamHalo} object that uses our stream-triggered API.
First, it finds the stream object associated with the current Kokkos execution space and uses it to initialize an \MPISX{Queue}; this ties progress for communication operations on this queue to the associated GPU stream.
Next, using \MPISX{Send\_init} and \MPISX{Recv\_init} calls, the constructor creates persistent send and receive requests for performing communication associated with \texttt{StreamHalo} gather operations.
Finally, the \texttt{StreamHalo} constructor performs an \MPISX{Matchall} operation on the created persistent operations to enable their use on stream. This call is a key difference between our API and the traditional MPI persistent communication API, and it moves message matching out of the communication critical path. 

Fig.~\ref{fig:enqueuegather} shows the code for the  \texttt{enqueueGather} method using the proposed stream-triggered API. As is common in GPU stream code, this code enqueues a sequence of operations to the GPU stream and associated \MPISX{Queue}. In order, the enqueued opertions (1) start the previously initialized and matched MPI receive requests, (2) pack communication buffers for sending, (3) start the previously initialized send requests, (4) wait for pending sends and receives to complete, and (5) unpack received communication buffers. 

With the exception of the lack of fencing, the ability to enqueue MPI operations to GPU streams, and \MPISX{Matchall}, this version has only minor changes compared to the standard MPI code. 
The API preserves ease of use, while enabling increases in efficiency and speed.
\section{Achieving CPU-Free MPI GPU Communication}
\label{sec:implementation}

In this section, we describe the design of a high-performance implementation of the communication of the API described in Sections~\ref{sec:api} and~\ref{sec:example} for the HPE Slingshot 11 NIC. We first describe the general data movement strategy, and then discuss a key issue for CPU-free two-sided MPI communication: delaying transmission on the sender until the remote receive buffer is ready, along with potential sources of deadlock in this implementation.

\subsection{Send and Receive with GPU Communication Counters}

\begin{figure*}
\begin{subfigure}[t]{\columnwidth}
\includegraphics[width=\textwidth]{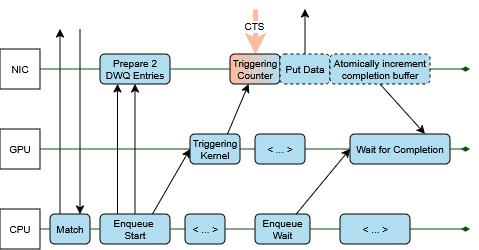}
\caption{Send timeline}
\label{fig:DWQTimeline:send}
\end{subfigure}
\begin{subfigure}[t]{\columnwidth}
\includegraphics[width=\textwidth]{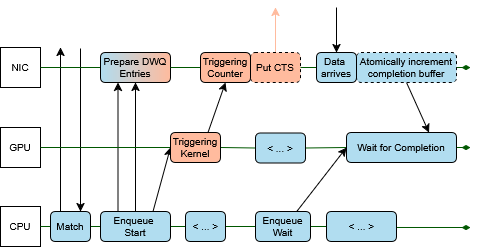}
\caption{Receive timeline}
\label{fig:DWQTimeline:receive}
\end{subfigure}
\caption{Timeline for our stream-triggered implementation of send, ready send, and receive operations with respect to the CPU, GPU, and NIC. Dotted boxes represent deferred work queue operations being carried out; orange boxes represent actions specific to the implementation of regular (non-ready) send, as described in Section~\ref{sec:impl:readiness}.}
\label{fig:DWQTimeline}
\end{figure*}

On creation of a persistent send and receive, our implementation allocates the following resources:
\begin{itemize}
    \item A triggering counter on the sender that can be triggered by the GPU stream to initiate communication.
    \item A local completion buffer for the persistent operation that can be polled by a GPU stream or kernel operation.
    \item A completion counter on both the sender and receiver that is used to trigger an \texttt{fi\_atomic} to increment the local completion buffer. On the sender, this counter increments on completion of the triggered \texttt{fi\_write}. On the receiver, this counter increments on an \texttt{FI\_REMOTE\_WRITE} to the memory buffer associated with the receive.
\end{itemize}

Our overall data-movement strategy for matching and ready-send/receive operations using these resources is shown in blue in Fig.~\ref{fig:DWQTimeline}:
\begin{itemize}
    \item \textbf{\MPISX{Match} operations on persistent sends and receives} use standard MPI communication operations to exchange RMA communication keys with all peers with which they are matching.
    \item \textbf{\MPISX{Enqueue\_Start} on a send operation} constructs an \texttt{fi\_write} deferred work queue entry for the data movement operation and an \texttt{fi\_atomic} deferred work queue entry to a \emph{local} memory completion buffer. The \texttt{fi\_write} deferred work queue entry is tied to the triggering counter that the local GPU will increment, and the \texttt{fi\_atomic} deferred work queue entry is tied to the completion counter for the \texttt{fi\_write} operation. Finally, \MPIS{Enqueue\_Start} on a send enqueues a GPU kernel that will increment the triggering counter associated with the persistent stream-triggered send. As shown in Fig.~\ref{fig:DWQTimeline:send}, when this kernel runs, it will trigger the deferred \texttt{fi\_write} operation, and the completion of this \texttt{fi\_write} will trigger the \texttt{fi\_atomic} operation to update the local completion buffer associated with the persistent send.
    \item \textbf{\MPIS{Enqueue\_Start} on a receive operation} creates a deferred work queue entry for an \texttt{fi\_atomic} operation that triggers  from the completion counter for \texttt{FI\_REMOTE\_WRITE} operations to the local memory buffer. As shown in Fig.~\ref{fig:DWQTimeline:receive}, when the \texttt{fi\_write} associated with a send places data in this buffer, the local \texttt{fi\_atomic} operation will update the local completion buffer associated with the persistent receive.
    \item \textbf{\MPISX{Enqueue\_Wait}} enqueues a kernel that polls for the relevant local completion buffer to be incremented to the appropriate value by an \texttt{fi\_atomic} operation. 
\end{itemize}
Finally, when possible, our implementation merges the use of triggering counters. Specifically, \MPISX{Enqueue\_startall} for ready sends and receives can share a triggering counter. Because non-ready sends require more involved synchronization, as described below, they cannot share triggering counters.

\subsection{Handling Receiver Readiness Checks}
\label{sec:impl:readiness}
Unlike many one-sided operations or MPI ready sends, standard MPI send operations cannot deliver data into the user's receive buffer until a matching receive operation is called. Because our current implementation relies on one-sided \texttt{fi\_write} operations to move data directly to the user buffer, it must include a mechanism to defer writes until a receive is triggered by the GPU on the receiver. 

We implement these semantics using the operations shown in orange in Fig.~\ref{fig:DWQTimeline}. First, \MPISX{Enqueue\_Start} on a receive matched with a regular non-ready send also creates a deferred \texttt{fi\_atomic} operation to a buffer to serve as a Clear-To-Send (CTS) message, and enqueues a kernel to trigger this operation when GPU execution reaches the receive. Second, regular sends tie counting of \texttt{FI\_REMOTE\_WRITE} to this buffer to its send triggering counter. Finally, enqueued regular sends only trigger on \emph{even numbered} counter values. Because MPI forbids starting another send or receive without an intervening wait, this guarantees that a stream-triggered regular send will only occur when (1) the GPU has triggered the relevant send and (2) a CTS for the relevant send has been received.

\subsection{Deadlock and Resource Exhaustion Analysis}
Because of their deferred execution model, understanding and limiting potential sources of deadlock are important when designing and implementing GPU-triggered communication. Our API and implementation are designed to minimize the potential for communication-induced deadlock by allowing \emph{only} \MPISX{Enqueue\_wait} to block stream execution; \MPISX{Enqueue\_start} cannot enqueue any GPU kernels that can block. This places the onus on MPI programmers to guarantee that all relevant sends and receives are active prior to a wait occurring, but MPI already requires this. 

Our current implementation includes another potential source of deadlock: resource exhaustion of Slingshot DWQ entries. Because calls on the CPU to \MPISX{Enqueue\_start} allocate DWQ entries, exhausting the pool of available entries can cause \MPISX{Enqueue\_start} to hold on the CPU while waiting for an arbitrary communication operation to complete and free an entry. Current Slingshot NICs have approximately 500 DWQ entries available for application use, and our implementation uses at most 2 DWQ entries per send or receive, so this has not been a significant limitation in practice thus far. More generally, an implementation could use a CPU thread to progress stream-triggered operations when hardware DWQ entries are exhausted.

Finally, we note that, like all GPU communication approaches, our proposed stream-triggering API relies on users to avoid deadlocks induced by the GPU scheduler failing to schedule streams with communication operations. The most common approach to doing this, for example when using NCCL in PyTorch, is to perform all communication on a high-priority GPU stream.

\section{Evaluation}
\label{sec:evaluation}

We measured the performance of the co-designed API and its prototype implementation described in the previous sections to evaluate its strengths and weaknesses in comparison with the standard MPI implementations. Here, we describe the benchmarks and  platforms used for this evaluation. We used microbenchmarks of point-to-point latency and bandwidth, and we measured the impact of our proposed API and implementation on the scalability of halo-exchange communications.

\subsection{Experimental Setup}
\label{sec:eval:setup}

We conducted our evaluation using two  benchmarks that vary in their behavior on two  platforms, which we describe below. Full implementation and benchmark source code, system run scripts, measured data, and analysis are available as open source online (Section~\ref{sec:artifact}).

\subsubsection{Test Systems}
We performed our tests on two systems: ORNL Frontier and LLNL Tuolumne. Frontier is a 9,408 node Cray EX system with one 64-core AMD EPYC CPU and four AMD Instinct MI250X GPUs per node. Each MI250X has two separately programmable 64GB graphics compute dies (GCDs), so applications can use up to eight GCDs per Frontier compute node. Tuolumne is a 1,152 node Cray EX supercomputer with four MI300A APUs per node, each with 24 AMD EPYC CPU cores and one GPU that share 128GB of high-bandwidth memory. Both Frontier and Tuolumne nodes are connected via a 200Gbps HPE Slingshot 11 network using a Dragonfly network topology, with four Slingshot NICs per node. Software versions used on these systems are shown in Table~\ref{tab:eval:setup:software}.
\begin{table}[t!]
    \centering
    \caption{Software versions used on test systems}
    \begin{tabular}{|c|r|r|r|r|}
    \hline
    System & Cray Compiler & ROCm & Cray  & Libfabric  \\
           & Environment   &      & MPICH &            \\
    \hline\hline
    Frontier & 18.0.1 & 6.2.4 & 8.1.31 & 1.22.0 \\
    \hline
    Tuolumne & 20.0.0 & 6.4.0 & 9.0.1 & 1.15.2  \\
    \hline
    \end{tabular}
    \label{tab:eval:setup:software}
\end{table}

\subsubsection{GPU Ping-Pong Microbenchmark}

To examine the fine-grain performance differences between communication using standard GPU-aware \texttt{MPI\_Send} and \texttt{MPI\_Recv} operations and our proposed stream-triggered alternatives, we designed a simple GPU-focused ping-pong benchmark that repeatedly packs, sends, receives, and unpacks a GPU buffer of configurable size between a pair of MPI processes a configurable number of times. For standard MPI operations, this GPU ping-pong adds two kernel launches and one GPU synchronization to the communication fast path compared to normal MPI ping-pong benchmarks: launching a packing kernel synchronizing the CPU with the GPU prior to calling \texttt{MPI\_Send}, and launching an unpacking kernel after the completion of the \texttt{MPI\_Recv}. In contrast, all packing, unpacking, and communication operations for the ping-pong exchange are enqueued to the GPU stream and associated \MPISX{Queue} at benchmark startup, with the CPU using only a single GPU/CPU synchronization to wait for all message exchanges to finish and time the length of the entire exchange. 

After completing warm-up iterations, we ran each ping-pong trial for a given buffer size for a number of round-trip iterations chosen to make the total measurement time between 1 and 30 seconds; this was 100,000 iterations for buffer sizes less than 4MB, 10,000 iterations for buffer sizes between 4MB and 64MB, and 1,000 iterations for buffer sizes greater than 64MB. We ran five ping-pong trials for all power-of-two buffer sizes between one byte and one gigabyte, and computed and plotted the average and 95\% confidence interval of the per-trial average ping-pong latency and bandwidth. 

\subsubsection{CabanaGhost Halo Exchange Benchmark}

To measure the impact of stream triggering on more complex communication patterns, we developed a C++-based halo exchange benchmark called \texttt{CabanaGhost} using the Cabana performance portability framework~\cite{slattery_cabana_2022}. \texttt{CabanaGhost}'s Game of Life benchmark uses an underlying Cabana grid halo exchange class to perform 8-point stencil communications, and we added a \texttt{StreamHalo} class to Cabana to support stream-triggered halo exchanges in \texttt{CabanaGhost}, as outlined in Section~\ref{sec:example}. The \texttt{StreamHalo} class can be configured to use standard MPI sends/receives with GPU fence operations, the stream-triggered standard sends and receives, or stream-triggered ready sends and receives.

We tested the performance impact of stream-triggered communication on \texttt{CabanaGhost} halo exchanges by examining the strong scaling performance of the Game of Life test on small and large problems. The large problem, a 30GB problem on the  Frontier system and a 62GB problem on the Tuolumne system, was chosen to test strong scaling performance below and up to the strong-scaling limit of the problem. The small problem, a 2GB problem on both systems, was chosen to test both regular and stream-triggered performance \emph{beyond} the strong scaling limit of the problem. For each we conducted five trials of 1,000 Game of Life iterations, where we measured the time from the beginning to the end of the solver phases of the benchmark.

\subsection{Ping-Pong Latency and Bandwidth}
\label{sec:eval:ping-pong}

\begin{figure}
\begin{subfigure}{\columnwidth}
\includegraphics[width=0.9\textwidth]{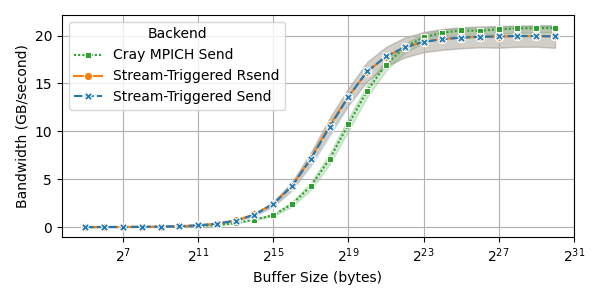}
\caption{Bandwidth}
\end{subfigure}
\begin{subfigure}{\columnwidth}
\includegraphics[width=0.9\textwidth]{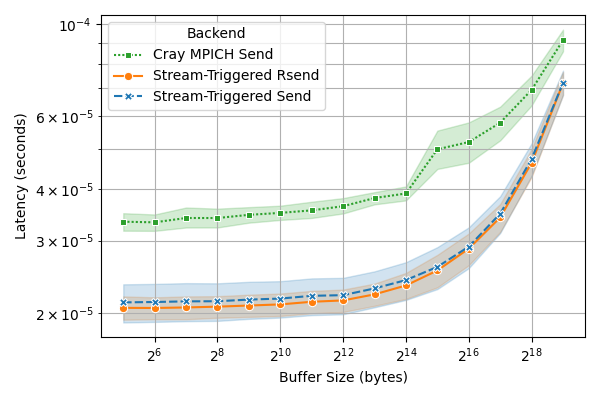}
\caption{Latency}
\end{subfigure}
\caption{Cray MPICH and Stream-Triggered GPU packing ping-pong bandwidth and latency between two Frontier nodes.}
\label{fig:pingpong}
\end{figure}
We first examined the ping-pong bandwidth and latency performance of our stream-triggered operations in comparison to Cray MPICH on the Frontier system, as shown in Fig.~\ref{fig:pingpong}. These results show that stream-triggered exchanges have significantly lower latency than CPU-triggered sends and receives, as well as significantly higher ping-pong bandwidth, particularly for messages between 4,096 bytes and one megabyte, but that Cray MPICH achieves higher bandwidth than our stream-triggered interface for messages 8MB or larger. On Frontier, stream-triggered sends and ready sends, respectively, have 12--39\% and 12--49\% lower latency than Cray MPICH for message sizes between 32 bytes and 512KB, and 2.7--4.9\% increased latency for messages 8MB and larger. On Tuolumne, stream-triggered sends and ready sends, respectively, have 14--51\% and 14--59\% lower latency than Cray MPICH for message sizes between 32 bytes and 2MB, and 8.6--18\% increased latency for messages 16MB and larger.

We believe the reduced latency of stream-triggered messages results from the elimination of kernel launch overheads and GPU/CPU synchronization from the communication critical path. However, this is a best-case improvement for stream-triggering latency because it requires two kernel launches and a synchronization per send; more complex communication patterns like  halo exchanges 
will incur these overheads only once per halo exchange. In addition, the ping-pong benchmark generally guarantees that the receiving buffer will be ready, minimizing the need for regular sends to wait for CTS arrival.

\subsection{Halo Exchange Strong Scaling}
\label{sec:eval:halo}
We tested strong-scaling \texttt{CabanaGhost} problems on power-of-two numbers of nodes from 1 to 1,024 nodes on both the ORNL Frontier and LLNL Tuolumne systems, using either 1, 2, 4, or 8 MI250X GPU GCDs (and MPI processes) per node on Frontier and 1, 2, or 4 MI300A APU GPUs on Tuolumne. On both systems, we ran five independent runs of 1,000 \texttt{CabanaGhost} iterations for each node/GPUs per node combination. On both systems, our evaluation focused on the average speedup over the mean single-GPU Cray MPICH solve time averaged across all samples with the same number of total MPI ranks.

\subsubsection{ORNL Frontier}
\label{sec:eval:halo:frontier}

\begin{figure*}
\begin{subfigure}{\columnwidth}
\includegraphics[width=\textwidth]{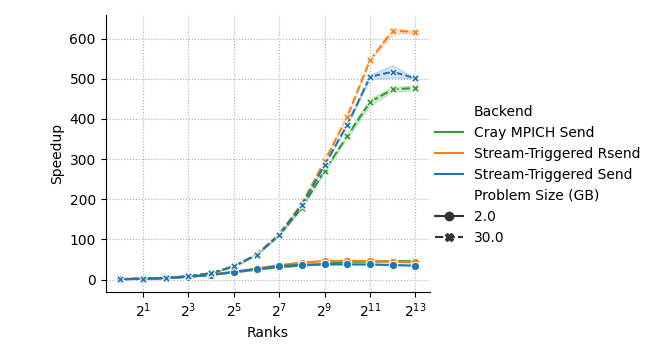}
\caption{Linear Scale}
\end{subfigure}
\begin{subfigure}{\columnwidth}
\includegraphics[width=\textwidth]{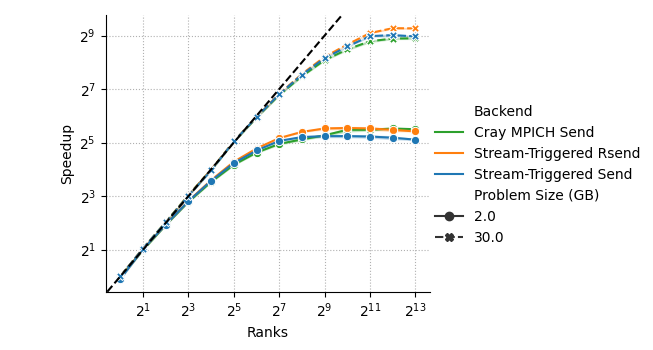}
\caption{Log Scale}
\end{subfigure}
\caption{Strong scaling speedup of the 8-point halo exchange benchmark for small and large problems on ORNL Frontier.}
\label{fig:halo:speedup:frontier}
\end{figure*}
Fig.~\ref{fig:halo:speedup:frontier} shows the results of this test for Frontier using both a log and linear scale, with the 95\% confidence interval of the speedup shown in the shaded area. As a baseline, the average Frontier solve time for the 30.0 GB problem using one GCD with the Cray MPICH backend was approximately 88.67 seconds, while the average solve time for the 2.0 GB problem using one GCD was 6.156 seconds.

Stream-triggered ready send has a higher maximum speedup than both stream-triggered standard send and Cray MPICH standard sends on both problems. On the large problem, both stream-triggered alternatives outperform Cray MPICH standard sends. On the small problem, stream-triggered standard send outperforms Cray MPICH up to 256 MPI processes, but Cray MPICH standard sends outperforms stream-triggered standard sends at 512 MPI processes and larger.

Table \ref{tab:speedup-frontier-peak} summarizes the maximum mean speedup for each backend, and the number of nodes and MPI processes at which this maximum is achieved.
\begin{table}[b]
    \centering
    \caption{Summary of best Frontier mean speedup by backend and problem size, along with the the number of nodes and MPI processes at which the maximum was reached. }
    \begin{tabular}{|r|r|r|r|}
    \hline
    Problem     & Backend     & Maximum & Nodes/Ranks \\ 
    Size (GB)   &             & Mean Speedup      &   \\
    \hline
    2.0         & Cray MPICH  &   50.23   & 1,024/1,024 \\
    \hline
                & Stream Rsend&   50.23   &  512/1,024  \\
    \hline
                & Stream Send &  40.49   &   512/1,024  \\
    \hline\hline
    30.0        & Cray MPICH &   485.4   &   1024/4096 \\
    \hline
                & Stream Rsend & 622.2   & 1,024/8,192 \\
    \hline
                & Stream Send &  543.7   & 1,024/4,096 \\
    \hline
    \end{tabular}
    \label{tab:speedup-frontier-peak}
\end{table}
Stream-triggered ready send achieves identical speedup to Cray MPICH on half as many nodes on the small problem and has a 28\% higher maximum mean speedup on the large problem.

\subsubsection{LLNL Tuolumne}
\label{sec:eval:halo:tuolumne}


\begin{figure}
\includegraphics[width=\columnwidth]{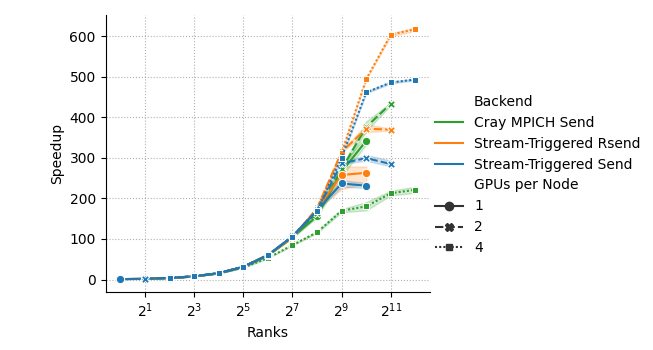}
\caption{Strong scaling speedup of the 8-point halo exchange benchmark on LLNL Tuolumne for the 62.0 GB problem with varying numbers of GPUs per node.}
\label{fig:halo:speedup:ppn:tuolumne}  
\end{figure}
Fig.~\ref{fig:halo:speedup:ppn:tuolumne} shows the result of the halo exchange strong scaling test for Tuolumne, using a linear scale with the 95\% confidence interval of the speedup shown in the shaded area. Unlike Fig.~\ref{fig:halo:speedup:frontier}, this chart plots speedup with different numbers of GPUs per node separately  due to large variation in speedup on Tuolumne with the same number of total GPUs but different numbers of GPUs per node. On this system, the average solve time for the 62.0 GB problem on one GPU with the Cray MPICH backend was approximately 88.80 seconds, while the average solve time for the 2.0 GB problem on one GPU was 3.110 seconds.

Broad Tuolumne performance trends are similar to Frontier, but with much higher variation in Cray MPICH performance with different numbers of GPUs per node. Specifically, 
(1) our stream-triggering operations are faster at all numbers of MPI processes when using four GPUs per node, (2) for a fixed number of total GPUs, Cray MPICH is always slower when using four GPUs per node, and (3) Cray MPICH outperforms stream-triggering at the largest scales when using one or two GPUs per node. We did not observe this behavior on Frontier, where speedup variance with different GCDs per node was almost always less than 5\%. We believe this may be because optimizations of Cray MPICH for Tuolumne MI300A APU features are still in progress.

\subsection{Communication Performance by Maximum Message Size}
\label{sec:eval:message-size}
As previously shown in Section~\ref{sec:eval:ping-pong}, our proposed stream-triggering primitives can significantly reduce the latency of individual message exchanges by reducing or removing kernel launch and GPU/CPU synchronization overheads. However, as Fig.~\ref{fig:DWQTimeline} shows, our simple approach to stream triggering, which always uses the NIC and does not currently attempt to optimize for different message sizes, incurs additional costs for regular (non-ready) sends. In contrast, production MPI implementations such as Cray MPICH use optimized hardware and protocols for different messages sizes. As a result, the performance benefits of our current stream-triggering implementation may vary significantly by message size, especially with more complex communication patterns.

\begin{figure}
\includegraphics[width=\columnwidth]{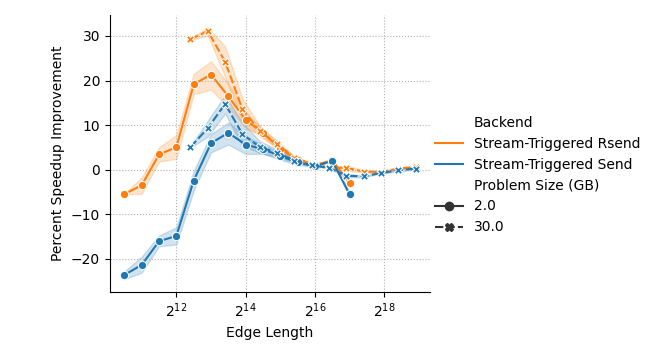}
\caption{Percentage improvement over Cray MPICH in the halo exchange benchmark speedup by problem size and average halo edge length on ORNL Frontier.}
\label{fig:halo:percent:edge}
\end{figure}
Fig.~\ref{fig:halo:percent:edge} plots the percentage speedup of the stream-triggering halo exchange approaches in comparison with Cray MPICH versus the average length of the edge in the halo exchanges on Frontier; note that four 8-byte corner messages are also exchanged each iteration in addition to these edges.

This graph shows that our stream-triggering implementation outperforms Cray MPICH for medium-size messages, which are performance critical in many HPC workloads. However, Cray MPICH significantly outperforms stream-triggered regular (non-ready) sends when messages are less than 8KB in size because Cray MPICH is able to use Slingshot unexpected message hardware to avoid waiting for CTS messages prior to sending messages. A stream-triggering implementation could also optimize such transfers using bounce buffers and GPU copies, but we have not yet implemented this optimization. Finally, we note that these costs were not captured in the micro-benchmark analysis in Section~\ref{sec:eval:ping-pong}, indicating the importance of assessing the performance of communication abstractions on more complex communication patterns.

\section{Related Work}
\label{sec:related}

Many MPI APIs for GPU communication have been proposed, as recently surveyed~\cite{Bridges:2025:Understanding}. Of these, the HPE two-sided GPU communication API~\cite{namashivayam_exploring_2022} and the MPICH \texttt{MPIX\_Stream} API~\cite{zhou_mpix_2022} are closest to our work, and key features from these proposals informed our design. Unlike them, our API leverages persistent communication to reduce the number of operations to add to the API and augments two-sided matching semantics to enable one-sided data movement.

Of the MPI GPU-triggered communication implementations, we are aware of only two that attempt to provide CPU-free communication: HPE's MPI one-sided communication API and implementation~\cite{namashivayam_exploring_2023}, and a GPU-initiated MPI partitioned communication system~\cite{temucin_design_2024}. The HPE one-sided API and implementation informed our design, particularly the use of \texttt{fi\_write} for data movement and chained  \texttt{fi\_atomic} opertions for completion notification. Our API and implementation provides similar data movement and notification capabilities while preserving two-sided communication semantics and enabling faster communication through ready sends. Unlike our approach, the GPU-initiated partitioned communication system in~\cite{temucin_design_2024} requires the use of CPU-based \texttt{MPI\_Pbuf\_prepare} calls to guarantee receiver readiness. Unlike our approach, this API is also very different from the standard MPI two-sided communication API.

Non-MPI APIs have also been designed and implemented that support CPU-free communication, particularly NVIDIA's NVSHMEM~\cite{nvidia_nvshmem_2025} and NCCL~\cite{nvidia_nccl_2025} communication systems. NCCL is most similar to our approach in providing an API for stream-initiated communication inspired by MPI; NVSHMEM is a one-sided communication API that relies on kernel-initiated communication. While the NCCL API is partially inspired by MPI, it provides a fundamentally collective communication API even for point-to-point communication, and, unlike our API,
does not support important MPI features such as message matching. Finally, NCCL has recently implemented CPU-free communication~\cite{hamidouche_gpu-initiated_2025}; we have not been able to compare our performance with this system because our API has not been ported to Infiniband, and NCCL GPU-Initiated Networking has not been ported to HPE Slingshot.
\section{Directions for Future Work}
\label{sec:future}

There is a wide range of directions for future work that builds on the research described in this paper. First, there are many optimization opportunities in the current implementation that have not yet been explored. For example, Section~\ref{sec:eval:message-size} shows that small stream-triggered sends when the receiver is not yet ready incur a substantial performance penalty. This case could be easily optimized for small messages using a simple bounce-buffer scheme, where small writes are staged through pre-allocated buffers associated with the persistent receive operation, and the \MPISX{Enqueue\_wait} kernel copies received data to its final destination. Similarly, the implementation could be optimized to leverage GPU-to-GPU copy hardware for intra-node communication.

Another potential area for future research is in stream-triggered collective communication operations. The OFI deferred work queue operations we leverage in this paper were originally designed for implementing offloaded collective communications~\cite{underwood_enabling_2011}, and they provide an opportunity for constructing flexible, high-performance stream-triggered operations. There are also many
tradeoffs with NIC, CPU, and GPU resource usage when implementing collectives of different buffer sizes and numbers of MPI processes.

Future work would also be helpful on the impact of this API and its implementation on a 
broader range of applications and NIC and GPU architectures. For example, integration of the API into additional applications such as machine learning and conjugate gradient solvers~\cite{trotter_cpu-_2025} would broaden the understanding of its impact. Testing our API on NVIDIA GPU/HPE Slingshot systems is in theory straightforward; our implementation relies only on GPU memory operations that are available on NVIDIA GPUs and should run with minimal changes on NVIDIA GPU/HPE Slingshot NIC systems. Unfortunately, the only such systems to which we have access did not include the NVIDIA kernel device driver option needed to support the required peer memory operations. Finally, porting our proposed API to NVIDIA InfiniBand NICs would enable systematic comparisons with NCCL GPU-Initiated Communication~\cite{hamidouche_gpu-initiated_2025}, insight into how well our API support for CPU-free communication on other network devices, and direct comparisons of the CPU-free communication features of HPE Slingshot and NVIDIA InfiniBand network devices.
\section{Conclusions}
\label{sec:conclusions}

This paper presented a co-designed API and implementation for MPI stream-triggered GPU communication that completely removes the CPU from the communication fast path. The API, which leverages persistent communication operations, preserves the matched two-sided communication semantics of MPI with only minimal changes; as a result, common HPC communication patterns such as halo exchanges remain easy to implement. The co-designed API and implementation leverage features of the OFI interface that the HPE Slingshot NIC makes available to GPU kernels, to provide complex synchronization without introducing additional blocking paths on GPU streams, while reducing small and medium message latency measured using a microbenchmark by up to 50\%. An evaluation of the impact of this API and implementation on a complex halo exchange benchmark running on up to 8,192 GPUs shows that they can increase the strong scaling speedup of such communication patterns by up to 28\%.
\section*{Source Code and Data Availability}\label{sec:artifact}

Source code for the stream-triggering primitives described in this paper is available as open source software as part of the MPI Advance family of libraries at \url{https://github.com/MPI-Advance/stream-triggering}. Source code for our modified version of the Cabana library is available at \url{https://github.com/CUP-ECS/Cabana}. Source code for the CabanaGhost benchmark used in this paper is available at \url{https://github.com/CUP-ECS/CabanaGhost}. Spack packages for building these libraries and benchmarks are available at \url{htts://github.com/CUP-ECS/spack-packages}. The run scripts, raw generated data, and analysis scripts used to create the figures found in this paper are available at \url{https://github.com/CUP-ECS/stream-triggering-data}.
\section*{Acknowledgements}

This research was supported in part by awards from the National Science Foundation under grant numbers 2103510, 2514054, 2450093, 2412182, and 2405142,  and by the U.S. Department of Energy's National Nuclear Security Administration (NNSA) under the Predictive Science Academic Alliance Program awards DE-NA0003966 and DE-NA0004267. 

This research used resources of the Oak Ridge Leadership Computing Facility at the Oak Ridge National Laboratory, which is supported by the Advanced Scientific Computing Research programs in the Office of Science of the U.S. Department of Energy under Contract No. DE-AC05-00OR22725.

Sandia National Laboratories is a multimission laboratory managed and operated by National Technology \& Engineering Solutions of Sandia, LLC, a wholly owned subsidiary of Honeywell International Inc., for the U.S. Department of Energy’s National Nuclear Security Administration under contract DE-NA0003525. This paper describes objective technical results and analysis. Any subjective views or opinions that might be expressed in the paper do not necessarily represent the views of the U.S. Department of Energy or the United States Government. SAND2026-17384O

\bibliographystyle{IEEEtran}
\bibliography{references.bib, software.bib}

\end{document}